\documentclass[12pt,fleqn]{article}
\usepackage{amssymb,graphicx,graphics,epsfig}
\usepackage{lscape,graphics,amsmath}
\usepackage{latexsym,amsfonts}
\usepackage[english]{babel}

\usepackage{epsfig}
\input epsf

\begin{document}
\begin{flushright}

\end{flushright}
%                          Title
\begin{center}
{\Large \bf Testing extra dimensions below the  production
threshold of  Kaluza-Klein excitations}

\vspace{4mm}

%                      author/address
Edward E.~Boos, Viacheslav E.~Bunichev,  Mikhail
N.~Smolyakov, Igor P.~Volobuev \\
\vspace{4mm}
 Skobeltsyn Institute of Nuclear Physics,
Moscow State University
\\ 119991 Moscow, Russia \\

\end{center}

\begin{abstract}
We consider a stabilized RS1 model in the energy range below the
direct production of  KK states. In this range we work out the
effective Lagrangian due to exchange of heavy KK tensor graviton
and scalar radion states and compute explicitly the  corresponding
effective coupling constants. As an example, the Drell-Yan lepton
pair production at the Tevatron and the LHC is analyzed in two
situations, when the first KK resonance is too heavy to be
directly detected at the colliders, and when the first KK
resonance is visible but other states are still too heavy. It is
shown that in both cases the contribution from the KK invisible
tower leads to a modification of final particles distributions. In
particular, for the second case a nontrivial interference between
the first KK mode and the rest KK tower takes place. Expected 95
\% CL limits for model parameters for the Tevatron and the LHC are
given. In the Appendix useful formulas for the cross sections and
distributions of various new $2\to2$ processes via heavy KK tower
exchange are presented, the new formulas containing nonzero
particle masses for final state fermions and bosons. The formulas
and numerical results are obtained by means of the CompHEP code,
in which all new effective interactions are implemented providing
a tool for simulation of  corresponding events and a more detailed
analysis.
\end{abstract}

\section{Introduction}
The paradigm of  modern quantum field theory implies that the
fundamental short-range interactions of elementary particles are
either  mediated by massive bosons or involve confinement. In
particular, the weak interactions are mediated  by massive gauge
bosons.

It is a common knowledge that historically the weak interactions
were first described by  contact four-fermion interactions,
because the large masses of the intermediate gauge bosons
disguised the nonlocal character of this interaction for small
energy or momentum transfer. Nevertheless, this approach enabled
the physicists to treat the weak interactions  theoretically long
before the consistent theory was formulated and the intermediate
gauge bosons were discovered.

Nowadays, as the high energy physics is looking for new
interactions beyond the Standard Model (SM), it is very likely
that we are in a situation, which is reminiscent of  Fermi's
theory of the weak interaction.

There are many theoretical schemes, which predict new interactions
beyond the Standard Model, mediated by new particles, but these
new particles may be too heavy to  be directly found in
experiments. Thus, it is worthwhile to consider the situation,
where the energies accessible at the existing and the upcoming
colliders are well below the threshold of production of these new
particles. In this case the new interactions, predicted by a
particular model, are reduced to contact interactions of the
Standard Model particles, which are defined by the model at hand.

The usage of contact interactions, or effective higher dimensional
operators, is a well-known way of parameterizing possible
deviations from the SM in a model independent way
\cite{Weinberg:1978kz,Buchmuller:1985jz,Burgess:1992gx,Burgess:1993vc}.
Such operators are  introduced by hand in phenomenological
extensions of the Standard Model, the only restriction on their
form usually  being the conservation of the Standard Model
symmetries.  In various processes at hadron and lepton colliders,
the effective operators may be probed with the aim to get some
first indications of a manifestation of physics beyond the SM or
to obtain restrictions on the parameters of the effective
Lagrangians (see, e.g.,
\cite{Whisnant:1997qu,Yang:1997iv,Boos:1999dd,Ferreira:2006in}).
However, there appears a large number of admissible effective
operators each coming with its own coupling constant, which
results in an uneasy problem of extracting  many parameters from
the experimental data.

Consideration of models with extra dimensions (see,
\cite{Antoniadis:1990ew,Antoniadis1993jp,Rizzo:1999pc,Carone:1999nz,
Davoudiasl:1999tf,Rizzo:1999en,Rizzo:1999br}) leads to a very
definite prediction for the structure of the contact interaction
operators entering the effective Lagrangian. In particular, the
contact interactions arising in such models are universal in the
sense that they are characterized by only one dimensional
constant. An experimental observation of such contact interactions
could be a strong  argument in favor of  models with extra
dimensions. Contact interactions due to summation of the exchange
of Kaluza-Klein (KK) towers within the ADD (Arkani-Hamed, Dimopoulos, Dvali) scenario were studied
in \cite{Hewett:1998sn}. The collider phenomenology of the contact
interactions appearing below the production threshold of KK modes
in the RS1 model,  such as changes of distribution tails, was
discussed in \cite{Davoudiasl:1999jd}. Contact interactions were
 also  considered in theories with warped universal extra dimensions,
where contributions of KK vector boson towers to Fermi's constant
were estimated \cite{Davoudiasl:1999tf}.

In this paper we  study the  stabilized RS1 (Randall, Sundrum 1) model below the
production threshold and perform a more accurate derivation of the
effective contact interaction Lagrangian. The latter enables us to
take into account the interactions of the scalar component of the
multidimensional gravity and to demonstrate explicitly that its
contributions to the contact interaction  are much smaller than
those of the tensor modes. From the analysis of Drell-Yan
distribution tails including KK contributions and those of the SM
with all the modern uncertainties and a natural restriction on the
KK resonance width to be smaller than its mass, we give expected
collider bounds on the effective Lagrangian parameters for both
the Tevatron and the LHC. For the case where the mass of the first
KK tensor mode lies within the collider energy reach we show how
the Breit-Wigner distribution of the resonance is modified due to
the contributions of all the remaining modes including the
destructive interference with the resonance.

The paper is organized as follows. First, within a stabilized
Randall-Sundrum model, we   derive  an effective Lagrangian for
the interactions of the Standard Model particles induced by
Kaluza-Klein excitations in the case, where the center of mass
energy is below the threshold of  the excitation production. An
important point is that we  explicitly calculate the effective
coupling constants. Next we discuss some collider manifestations
of the effective contact interactions giving the relevant formulas
in the Appendices. In particular, the formulas for the cross
sections and distributions include both tensor and scalar
contributions and take into account the masses of the final state
particles.

\section{The Effective Lagrangian for the stabilized RS1 model}

The characteristic feature of theories with compact extra
dimensions is the presence of towers of Kaluza-Klein excitations
of the bulk fields, all the excitations of a bulk field having the
same type of coupling to the fields of the Standard Model.  If we
consider such a theory  for the energy or momentum transfer much
smaller,  than the masses of the KK excitations, we can pass to
the effective "low-energy" theory, which can be obtained  by the
standard procedure. Namely, we have to drop the momentum
dependence in the propagators of the heavy modes and integrate
them out in the functional integral built with the  action of the
theory. This can be easily done, if the self-interaction of the
modes is weak, and one can drop it as well. As a result, we get a
certain contact interaction of the Standard Model fields for each
bulk field of the multidimensional theory. If we also assume that
the fundamental energy scale of the $(4+d)$-dimensional theory $M$
is of the same order of magnitude, as the inverse size of extra
dimensions, then  the masses of the KK excitations are
proportional to this energy scale  $M$, and the wave functions of
the modes are proportional to $M^{d/2}$. This defines the coupling
constant of the contact interactions up to a dimensionless factor.
The particular  structure of the  contact interaction Lagrangian
is fixed by the corresponding structure of the SM current coupled
to the zero mode of a bulk field  and by the spin-density matrix
of its KK modes. This leads to a number of very concrete
predictions for collider phenomenology.

  The bulk field, which appears in any theory with
extra dimensions, is the gravitational field. The question, what
fields, besides the gravitational one, propagate in extra
dimensions, has no unique answer yet. The theory of universal
extra dimensions \cite{Appelquist:2000nn} allows all the fields of
the Standard Model to propagate in the bulk. Other approaches
allow only some of the Standard Model fields to  propagate in
extra dimensions. In particular, another interesting assumption is
that only the gauge fields live in extra dimensions. A motivation
for this can be the theory of "fat" branes, where the fields of
the Standard Model are trapped on the brane by a particular
mechanism. It turns out that it is easy to trap the fermion fields
on the brane, but there is no mechanism yet for trapping the gauge
fields \cite{Rubakov:2001kp}.

There are two main approaches in theories with large extra
dimensions, -- the ADD scenario and the Randall-Sundrum model,
which admit the above discussed situation, where the
multidimensional Planck mass and the inverse size of the extra
dimensions are both in the $TeV$ energy range; the ADD scenario
\cite{ArkaniHamed:1998rs} in this case demands a very large number
of extra dimensions. Moreover, a flaw of this approach is the
assumption that  the multidimensional background metric can be
taken to be flat, which means that the proper gravitational field
of the brane can be neglected. The validity of the results
obtained within this approach depends upon whether this
approximation is good or not so good. In any case, studying the
equations of motion for multidimensional  gravity interacting with
a brane of non-zero tension, it is not difficult to understand
that if extra dimensions are compact, there should exist at least
two branes, and the background metric must be essentially
nonflat.

This situation is realized  in the Randall-Sundrum model with two
branes \cite{Randall:1999ee}, -- the RS1 model, which is one of
the most interesting brane world models. It is a consistent model
based on an exact solution for gravity interacting with two branes
in five-dimensional space-time. If our world is located on the
negative tension brane, it is possible to explain the weakness of
the gravitational interaction by the warp factor in the metric. A
flaw of this model is the presence of a massless scalar mode, --
the radion, which describes fluctuations of the branes with
respect to each other. As a consequence, one gets a
scalar-tensor theory of gravity on the branes, the scalar
component being described by the radion. It turns out that the
coupling of the massless radion to matter on the negative tension
brane contradicts the existing restrictions on the scalar
component of the gravitational interaction, and in order for the model to 
be phenomenologically acceptable the radion must acquire a mass.
The latter is equivalent to the stabilization of the brane
separation distance, i.e. it must be defined by the model
parameters. The models, where the interbrane distance is fixed in
this way, are called stabilized models, unlike the unstabilized
models, where the interbrane distance can be arbitrary. Below we
will discuss the contact interactions of the Standard Model
particles, which arise in the stabilized brane world model
proposed in \cite{DeWolfe:1999cp}.

Let us denote the coordinates  in five-dimensional space-time $E=M_4\times
S^{1}/Z_{2}$  by $\{ x^M\} \equiv \{x^{\mu},y\}$, $M= 0,1,2,3,4, \,
\mu=0,1,2,3 $, the coordinate $x^4 \equiv y, \quad -L\leq y \leq L$
parameterizing the fifth dimension. It forms the orbifold, which is realized
as the circle of the circumference $2L$ with the points $y$ and $-y$
identified. Correspondingly, the metric $g_{MN}$  and the scalar field
$\phi$ satisfy the orbifold symmetry conditions
\begin{eqnarray}
\label{orbifoldsym}
 g_{\mu \nu}(x,- y)=  g_{\mu \nu}(x,  y), \quad
  g_{\mu 4}(x,- y)= - g_{\mu 4}(x,  y), \\ \nonumber
   g_{44}(x,- y)=  g_{44}(x,  y), \quad
   \phi(x,- y)=  \phi(x,  y).
\end{eqnarray}
The branes are located at the fixed points of the orbifold, $y=0$
and $y=y_b= L$.

The  action of the stabilized brane world  model can be written as
\begin{eqnarray}\label{actionDW}
S&=& -2 M^3\int d^{4}x \int_{-L}^L dy  R\sqrt{-g}+ \int d^{4}x
\int_{-L}^L dy
\left(\frac{1}{2}g^{MN}\partial_M\phi\partial_N\phi-V(\phi)\right)\sqrt{-g}
-\\\nonumber & -&\int_{y=0} \sqrt{-\tilde g}\lambda_1(\phi) d^{4}x
+\int_{y=L}\sqrt{-\tilde g}(-\lambda_2(\phi) + L_{SM})   d^{4}x .
\end{eqnarray}
Here $V(\phi)$ is a bulk scalar field potential and
$\lambda_{1,2}(\phi)$ are quadratic brane scalar field potentials,
$\tilde{g}=det\tilde g_{\mu\nu}$, and $\tilde g_{\mu\nu}$ denotes
the metric induced on the branes.  The fields of the Standard
Model are assumed to be located on the negative tension brane,
$L_{SM}$ denoting the Standard Model Lagrangian. A background
solution in this theory, which preserves the Poincar\'e invariance
in any four-dimensional subspace $y=const$, looks like
\begin{eqnarray}\label{metricDW}
ds^2&=&  e^{-2A(y)}\eta_{\mu\nu}
{dx^\mu  dx^\nu} -  dy^2 \equiv\gamma_{MN}(y)dx^M dx^N, \\
\phi(x,y) &=& \phi(y),
\end{eqnarray}
the fields of the Standard Model being in the Higgs
vacuum; it is also assumed that the  vacuum  value of the Higgs potential is
equal to zero. For a special  choice of the potentials
\cite{DeWolfe:1999cp}, the solutions for  functions $A(y), \phi(y)$
are
\begin{eqnarray}\label{sols}
\phi(y) &=& \phi_1 e^{-u|y|}, \\ \nonumber
A(y) &=& k(|y|-L) + \frac{\phi_1^2}{48 M^3}(e^{-2u|y|} - e^{-2uL} ).
\end{eqnarray}
The solution for $A(y)$ is normalized so that the induced metric
on the brane at $y=L$ is flat and the coordinates $\{x^\mu\}$ are
Galilean on this brane \cite{LL,Rubakov:2001kp,Boos:2002ik}. The
constants $k, u$,  the boundary values of the scalar field
$\phi_{1,2}$, the fundamental five-dimensional energy scale $M$,
on a par with the coefficients of the quadratic brane potentials
$\lambda_{1,2}(\phi)$, are the parameters of the model. When the
former parameters are made dimensionless with the help of $M$,
they should be of the order $O(1)$, so that there is no
hierarchical difference in the parameters. The separation distance
in defined by the equation
\begin{equation}\label{sepL}
L= \frac{1}{u}\ln \left(\frac{\phi_1}{\phi_2}\right)
\end{equation}
and,  therefore,  it is stabilized.

As we have explained above, it is sufficient to treat the
gravitational interaction perturbatively to the linear order. To
this end we represent the metric and the scalar field as
\begin{eqnarray}\label{metricparDW}
g_{MN}(x,y)&=& \gamma_{MN}(y) + \frac{1}{\sqrt{2M^3}} h_{MN}(x,y), \\
\phi(x,y) &=& \phi_1 e^{-u|y|} + \frac{1}{\sqrt{2M^3}} f(x,y).
\end{eqnarray}
Substituting this representation into action (\ref{actionDW}) and
keeping the terms up to the second order in $1/\sqrt{2M^3}$, we
get the so called second variation  Lagrangian \cite{Boos:2005dc},
which includes the Fierz-Pauli Lagrangian in the warped
background,  the terms describing the interaction of the
linearized gravity in the background (\ref{sols}) with the branes,
and an interaction Lagrangian determining the coupling to the
fields of the Standard Model. In particular, the interaction
Lagrangian is
\begin{equation}\label{Lagr_int}
L_{int} = -\frac{1}{2\sqrt{2M^3}}h^{\mu\nu} T_{\mu\nu},\quad
T_{\mu\nu} = 2 \frac{\delta L_{SM}}{\delta \gamma^{\mu\nu}} -
\gamma_{\mu\nu} L_{SM} ,
\end{equation}
the energy-momentum tensor $T^{\mu\nu}$ being canonically built
from the Standard Model Lagrangian $ L_{SM}$.

In paper \cite{Boos:2005dc} it was shown that the action built
with the second variation  Lagrangian can be diagonalized for
any background solution and after the mode decomposition, which
includes integration over $dy$, brought to the form
\begin{equation}\label{redact}
S_{eff}=\frac{1}{4}\sum_{k=0}^\infty\int dx \left(
\partial^\sigma b^{k,\mu \nu}\partial_\sigma b_{\mu \nu}^k-m_k^2
b^{k,\mu\nu}b_{\mu \nu}^k \right)
+\frac{1}{2}\sum_{k=1}^\infty\int dx \left(\partial_\nu
\varphi_k\partial^\nu \varphi_k-\mu_k^2\varphi_k \varphi_k\right),
\end{equation}
with $m_0 = 0$ and the other  masses of the four-dimensional
tensor and scalar fields being defined by the background solutions
$A(y)$ and $\phi(y)$ and the model parameters. The free theory
with this Lagrangian can be easily quantized, the propagators of
the massive tensor and scalar particles are given by
\begin{eqnarray}\label{propagator_t}
D^k_{\mu\nu,\rho\sigma}(p)&=&
\frac{B^k_{\mu\nu,\rho\sigma}(p)}{p^2-m_k^2 + i\epsilon}\,\,,
\\\nonumber B^k_{\mu\nu,\rho\sigma}(p)&=& \frac{1}{2}\left(\eta_{\mu\rho}
-\frac{p_\mu p_\rho}{m_k^2} \right) \left(\eta_{\nu\sigma}
-\frac{p_\nu p_\sigma}{m_k^2} \right) +
\frac{1}{2}\left(\eta_{\mu\sigma} -\frac{p_\mu p_\sigma}{m_k^2}
\right) \left(\eta_{\nu\rho} -\frac{p_\nu p_\rho}{m_k^2} \right) -
\\\nonumber && - \frac{1}{3}\left(\eta_{\mu\nu} -\frac{p_\mu
p_\nu}{m_k^2} \right)
\left(\eta_{\rho\sigma} -\frac{p_\rho p_\sigma}{m_k^2} \right), \\
\label{propagator_s}
 D^k(p) &=& \frac{1}{p^2-\mu_k^2 + i\epsilon}\,\,.
\end{eqnarray}

The interaction of these four-dimensional fields with the fields of the
Standard Model is defined by the interaction Lagrangian (\ref{Lagr_int}) and
looks like
\begin{equation}\label{allint}
L_{int} = -\frac{1}{2\sqrt{2M^3}}\left(\psi_0(L)b_
{\mu\nu}^0(x)T^{\mu\nu}+\sum_{n=1}^\infty\psi_n(L)b_{\mu\nu}^n(x)T^{\mu\nu}+
\frac{1}{2}\sum_{n=1}^\infty  g_n(L) \varphi_n(x) T_\mu^\mu
\right),
\end{equation}
$\psi_n(y)$ and $g_n(y)$ being the wave functions of the modes in the extra
dimension. Thus,  the couplings are defined by the values of the wave
functions on the brane. The latter are specified by the background solutions
$A(y)$ and $\phi(y)$ and the values of the model parameters.
Since the field $b_{\mu\nu}^0(x)$ describes the massless graviton, the
coupling constant of this field to matter on our negative tension brane must
coincide with the inverse Planck mass. The latter can be expressed in terms
of the model parameters as \cite{Boos:2004uc}
\begin{equation}\label{couplingDW}
M_{Pl}^2=\frac{M^3}{u}e^{-2c}(2b)^{-\frac{k}{u}}\left\{\gamma\left(\frac{k}
{u},2b\right) -\gamma\left(\frac{k}{u},2be^{-2uL}\right)\right\}, \quad
b=\frac{\phi_1^2}{48 M^3},
\end{equation}
$\gamma$ denoting the incomplete gamma function  and constant  $c
= -kL -be^{-2uL}$. For $u \to 0$  (in fact, in this limit the
background scalar field becomes constant, its fluctuations
decouple from those of the gravitational field and the model goes
to the unstabilized Randall-Sundrum model) this expression goes to
the well known relation between the energy scales for an observer
on the negative tension brane in the unstabilized RS1 model
\begin{equation}\label{rel2a}
M^2_{Pl}=  M^3\frac{e^{2 kL }-1}{ k},
\end{equation}
which allows for a solution to the hierarchy problem of the
gravitational interaction, if $M\simeq k \simeq 1 TeV$ and $kL
\simeq 35$ \cite{Rubakov:2001kp,Boos:2002ik}. The problem of
energy scales in the RS1 model was discussed in detail in the paper
\cite{Boos:2004uc}. In  this paper the parameter space of the
model was scanned and different scenarios for the fundamental
energy scale and the KK excitations scale were studied. In
particular, it was shown that it was possible to have the
fundamental five-dimensional energy scale $M$  of the order of 1--10
$TeV$ with the masses of the tensor and the scalar KK excitations
also being in the  same energy range. The present day experimental
data imply that this scenario is more likely, than the scenarios
with the light radion. In this case the interactions of the
Standard Model particles at the accessible energies due to the KK
excitations of the tensor and the scalar fields can be very well
approximated by a contact interaction, because we can drop the
momentum dependence in the propagators.

Integrating out the heavy tensor modes in  the sum of Lagrangians
(\ref{redact}), (\ref{allint}) induces the interaction
of the Standard Model fields of the form
\begin{eqnarray}\label{int_T}
L_{T}&=&\frac{1}{8M^3}\left(\sum_{n >0} \frac{\psi^2_n(L)}{
m_n^2}\right) T^{\mu\nu}\Delta_{\mu\nu, \rho\sigma}T^{
\rho\sigma},\\ \Delta_{\mu\nu, \rho\sigma}
&=&B^k_{\mu\nu,\rho\sigma}(p)|_{p=0}=
\frac{1}{2}\eta_{\mu\rho}\eta_{\nu\sigma} +
\frac{1}{2}\eta_{\mu\sigma}\eta_{\nu\rho}-\frac{1}{3}
\eta_{\mu\nu}\eta_{\rho\sigma},
\end{eqnarray}
whereas integrating out the scalar modes  induces the interaction
of the form
\begin{equation}\label{int_S}
L_{S}=\frac{1}{64M^3} \left(\sum_n \frac{g^2_n(L)}{\mu_n^2}
\right) T^{\mu}_{\mu} T^{\nu}_{\nu}.
\end{equation}

The coupling constants $\frac{1}{8M^3}\left(\sum
\frac{\psi^2_n(L)}{ m_n^2}\right)$ and $\frac{1}{64M^3}\left(\sum
\frac{g^2_n(L)}{\mu_n^2} \right)$ can be approximately estimated
in the model as follows. For the stabilized Randall-Sundrum model
presented in \cite{Boos:2005dc}, it was shown that it was more
convenient to use parameters $b= {\phi_1^2}/{48 M^3} $, $\tilde k
= k -  2bu$,  and $L$ instead of $\phi_{1,2}$ and $k$.  It was
also shown that for $uL \ll 1$ the metric of the  stabilized model
is similar to that of the unstabilized one with the inverse
anti-de Sitter radius $\tilde k$ instead of $k$, and it is
possible to find analytical solutions for the wave functions of
the tensor and scalar modes and their mass spectra.

In particular, the spectrum of the tensor excitations is defined
by $J_{1}\left(\frac{m_{n}}{\tilde k}\right)=0$  and in the
approximation of an infinitely hard brane potential $\lambda_2$ we
get $\psi_{n}|_{y=L}=-\sqrt{\tilde k}$ (see
\cite{Boos:2002ik,Boos:2005dc}). The sum over the tensor modes in
(\ref{int_T}) can be estimated to be $$ \frac{1}{8M^3} \sum_{n>0}
\frac{\psi^2_n(L)}{m_n^2} \approx \frac{2}{\pi^2 M^3\tilde k}
\sum_{n>0} \frac{1}{(1+4n)^2} \approx \frac{0.1246}{8M^3\tilde
k}\approx \frac{1.82}{\Lambda_{\pi}^{2}m_{1}^{2}}\,, $$ where we
have introduced the coupling constant $\Lambda_\pi$ of the first
KK resonance and its mass $m_1$: $$
\frac{1}{\Lambda_{\pi}}=-\frac{\psi_{1}(L)}{\sqrt{8M^{3}}},\quad
m_{1}=3.83\tilde k. $$ It is worth mentioning that the
contribution of the first KK resonance to this sum is exactly $
\frac{1}{\Lambda_{\pi}^{2}m_{1}^{2}} $. One can also see that
$\frac{1}{\Lambda_{\pi}}=-\frac{\psi_{n}(L)}{\sqrt{8M^{3}}}$ for
relatively small $n$.

In this parametrization, which is often used in the RS1 model, the
effective Lagrangian for this model takes the form
\begin{eqnarray}\label{tensor_intCRS}
L_{eff}&=&L_{T}+L_{S}=\frac{1.82}{\Lambda_{\pi}^{2}m_{1}^{2}}
T^{\mu\nu} \tilde \Delta_{\mu\nu, \rho\sigma}T^{ \rho\sigma},\\
\tilde \Delta_{\mu\nu, \rho\sigma} &=&
\frac{1}{2}\eta_{\mu\rho}\eta_{\nu\sigma}  +
\frac{1}{2}\eta_{\mu\sigma}\eta_{\nu\rho}-\left(\frac{1}{3}-\frac{\delta}{2}\right)
\eta_{\mu\nu}\eta_{\rho\sigma},
\end{eqnarray}
where $\delta$ stands for the contribution of the scalar modes and
will be calculated below.

For the scalar sector  the spectrum in this approximation is
defined by \cite{Boos:2005dc} (where we have substituted $m_{1}$
instead of  $\tilde k$)

  $$ \left(1+ \alpha + \frac{3.83\cdot u}{m_{1}}\right)J_{\alpha}\left(\frac{3.83\cdot\mu_n}{m_{1}}\right) - \frac{3.83\cdot\mu_n}{m_{1}}
J_{\alpha-1}\left(\frac{3.83\cdot\mu_n}{m_{1}}\right) =0 $$ with
$\alpha = \sqrt{\left(1+ \frac{3.83\cdot u}{m_{1}} \right)^2 +
{8}\frac{3.83^{2}bu^2}{{m_{1}}^2}}\approx 1.8$ and for the wave
functions we get
 $$
g_{n}|_{y=L}=4 \sqrt{\frac{2}{3}} \frac{\sqrt{bu^{2}
m_{1}}}{\sqrt{3.83}\mu_{n}\sqrt{1-\frac{8 b u^{2}}{
\mu_{n}^{2}}}}. $$ We note that with given parameters $m_1$,
$\Lambda_\pi$ and $L$, describing tensor sector of the model (one
should take $\tilde k L \approx 35$ for the hierarchy problem to
be solved), the spectrum of the scalar modes and their couplings
to matter are also defined by the parameters $u$ and $b$. Quite an
interesting feature of the massive modes (both tensor and scalar)
is that their masses and coupling constants in fact do not depend
on the size of the extra dimension $L$, at least for relatively
small $n$.

To estimate the corresponding sum over the scalar modes, we should
specify the model parameters. Let us suppose that the lowest
scalar mode, the radion, has the mass of the order of $2 TeV$.
Such a situation can be realized if $\Lambda_{\pi}\simeq 8 TeV$,
$m_{1}\simeq 3.83 TeV$ (correspondingly, $M\simeq 2 TeV$, $\tilde
k \sim 1TeV$), $u\simeq 0.003 TeV$, $bu^{2}\simeq 0.28 TeV^{2}$.
In this case the sum over the scalar modes in (\ref{int_S}) turns
out to be $$ \sum_n \frac{g^2_n(L)}{\mu_n^2} \approx
\frac{3.83}{m_{1}} (0.341 + 0.002) \approx \frac{1.314}{m_{1}}\,,
$$ where the first term corresponds to the contribution of the
radion. Correspondingly, we find $\delta\approx 0.7$.

As we have mentioned above, this interaction Lagrangian leads to
quite definite processes with the SM particles, which are
determined by the structure of the energy-momentum tensor
$T^{\mu\nu}$. The latter is a sum of the energy-momentum tensors
of the free SM fields and  of contributions from the interaction
terms, which are proportional to the SM coupling constants. The
energy-momentum tensors of the free SM fields are quadratic in the
fields and are explicitly given in  Appendix A.

One can easily see that for massless vector fields the trace of
the energy-momentum tensor vanishes, and the scalar degrees of
freedom do not contribute to the effective interaction. They can
contribute to the effective interaction, if one takes into account
the conformal anomaly of massless fields. The anomalous part of
the energy-momentum tensor turns out to be
$$\Delta{T}_{\mu\nu}=\frac{b(g)}{6g}\left(\eta_{\mu\nu}-
\frac{\partial_{\mu}\partial_{\nu}}{\Box}\right)F_{\rho\sigma}F^{\rho\sigma},
$$ which gives the well-known expression for the anomalous trace
of this tensor $$\Delta {T}^{\mu}_{\mu}=
\frac{b(g)}{2g}F_{\rho\sigma}F^{\rho\sigma}, $$ where $b(g)$ is
the beta function. The structure of this anomalous term in the
energy-momentum tensor is such that the interaction due to the
exchange of tensor particles (\ref{int_T}) vanishes, and only the
interaction due to the exchange of scalar particles (\ref{int_S})
remains. However, this interaction is rather suppressed compared
to the one due to the exchange  of tensor particles, because the
trace of the energy-momentum tensor is proportional to the
particle mass  that is much smaller than both $m_1$ and
$\Lambda_\pi$. And a possibility to observe the scalar component
of the effective interaction may be due to the Higgs-radion mixing
\cite{Giudice:2000av,Csaki:2000zn}.

Thus, in the lowest nonvanishing order in the SM coupling
constants the effective Lagrangian (\ref{tensor_intCRS}) is a sum
of four-particle effective operators (not only 4-fermions, but
also 2-fermions--2-vectors, 4-vector particles etc.). Experimental
observation of production processes following from the effective
Lagrangian (\ref{tensor_intCRS}) or restrictions on their
cross-sections allow one to estimate the multidimensional energy
scale $M$, provided one gets a theoretical estimate  for the
product of the  parameters $m_1$ and $\Lambda_\pi$ in
(\ref{tensor_intCRS}). Their ratio may be estimated from the fact
that the width of the first KK excitation must be smaller than its
mass.

\section{Two body processes with KK gravitons}

As was mentioned, the lowest order effective Lagrangian in the SM
couplings  contains a sum of various four-particle
(not only 4-fermions, but also
2-fermions--2-bosons, 4-bosons) effective operators, which are
gauge invariant with respect to the SM gauge group and lead to a
well defined phenomenology. The Lagrangian involves only three free
parameters $\Lambda_\pi$, $m_1$ and $\delta$, where $\Lambda_\pi$, $m_1$
 parameterize the common
overall coupling and $\delta$  parameterizes the relative
contribution of the scalar radion field (or fields as takes place
in the stabilized RS model). In this paper we shall  not present a
detailed phenomenology, but rather point out some interesting
aspects. In the leading order only the  neutral currents of the
same generation SM fields are involved. These new interactions do
not lead to additional decay modes. Possible new decays of the SM
particles from the effective Lagrangian  may only be present in
the next order in the SM couplings, when charged currents appear
in the SM energy-momentum tensor. Also new effective 4-particle
operators following from the SM energy-momentum tensor obviously
do not lead to flavor changing neutral currents. In the tree level
approximation there are several processes following from the
effective Lagrangian, which appear only at loop level in the SM
such as $gg\to l^{+}l^{-}$,  $gg\to ZZ (W^+W^-)$, $e^{+}e^{-}\to
gg$, $\gamma\gamma \to gg$ etc. In  Appendix B analytical
expressions for the total and differential cross sections for the
processes $gg\to l^{+}l^{-}$,  $gg\to ZZ (W^+W^-)$, $q\bar{q}\to
l^{+}l^{-}$, $q\bar{q}\to ZZ (W^+W^-)$, $e^{+}e^{-}\to f\bar{f}$,
$e^{+}e^{-}\to gg$, $\gamma\gamma \to f\bar{f}$, $\gamma\gamma \to
gg$ are presented. For completeness we keep nonzero masses of the
final state particles. In the case of massless fermions formulas for
the total and differential cross sections for the Drell-Yan
processes $gg\to l^{+}l^{-}$ and $q\bar{q}\to l^{+}l^{-}$
 are in complete agreement with \cite{Hewett:1998sn,
Gupta:1999iy,Cheung:1999wt}. Formulas (\ref{ggzz})-(\ref{uuzz})
for processes $gg\to ZZ$ and $q\bar{q}\to ZZ$ that  take into
account scalar KK modes, massive final states and the interference
with the SM amplitudes are presented here for the first time. In
the cases where colliding gluons produce massive final particles,
there is also a scalar radion contribution, which is proportional
to the parameter $\delta^2$ of the order of 1 and to the trace
anomaly coefficient $(b(g_s)/2g_s)^2$. We give this contribution
in formulas, although numerically it is about $100$ times smaller
than the corresponding tensor contribution.

Below we will perform numerical simulations for the Drell-Yan
process because this channel is most sensitive for new physics.
Detailed simulations for other channels will be made in a further
study.

Symbolic and numerical computations, including simulations of the
SM background in a thought experiment for Tevatron and LHC, have been performed
by means of the version of the CompHEP \cite{comphep} package
realized on the basis of the FORM \cite{form} symbolic program. The
Feynman rules following from the effective Lagrangian have been
implemented into this version of the CompHEP. Such an
implementation allows one to use the code for event generation and
to perform analysis in future more realistic studies.

\begin{figure*}
\begin{minipage}[t]{.49\linewidth}
\centering
\includegraphics[width=8.5cm,height=8.5cm]{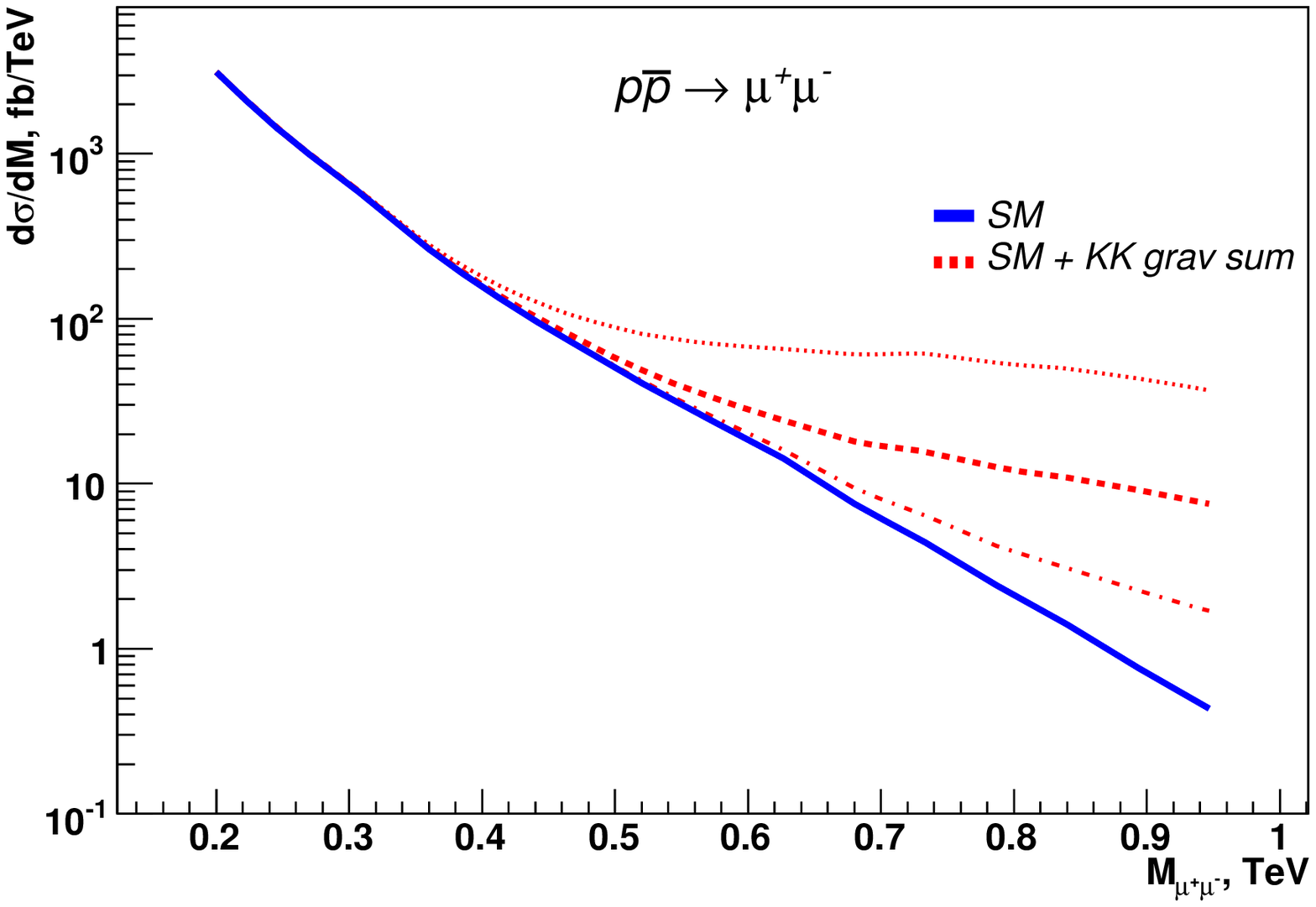}
\caption{Dilepton invariant mass distribution for  parameter
$\frac{0.91}{\Lambda_{\pi}^2 m_{1}^2}\times {TeV^4}$=0.66
(dashed-dotted line), 1.82 (dashed line), 4 (dotted line) for the
Tevatron} \label{KKcontrib1}
\end{minipage}
\begin{minipage}[t]{.49\linewidth}
\centering
\includegraphics[width=8.5cm,height=8.5cm]{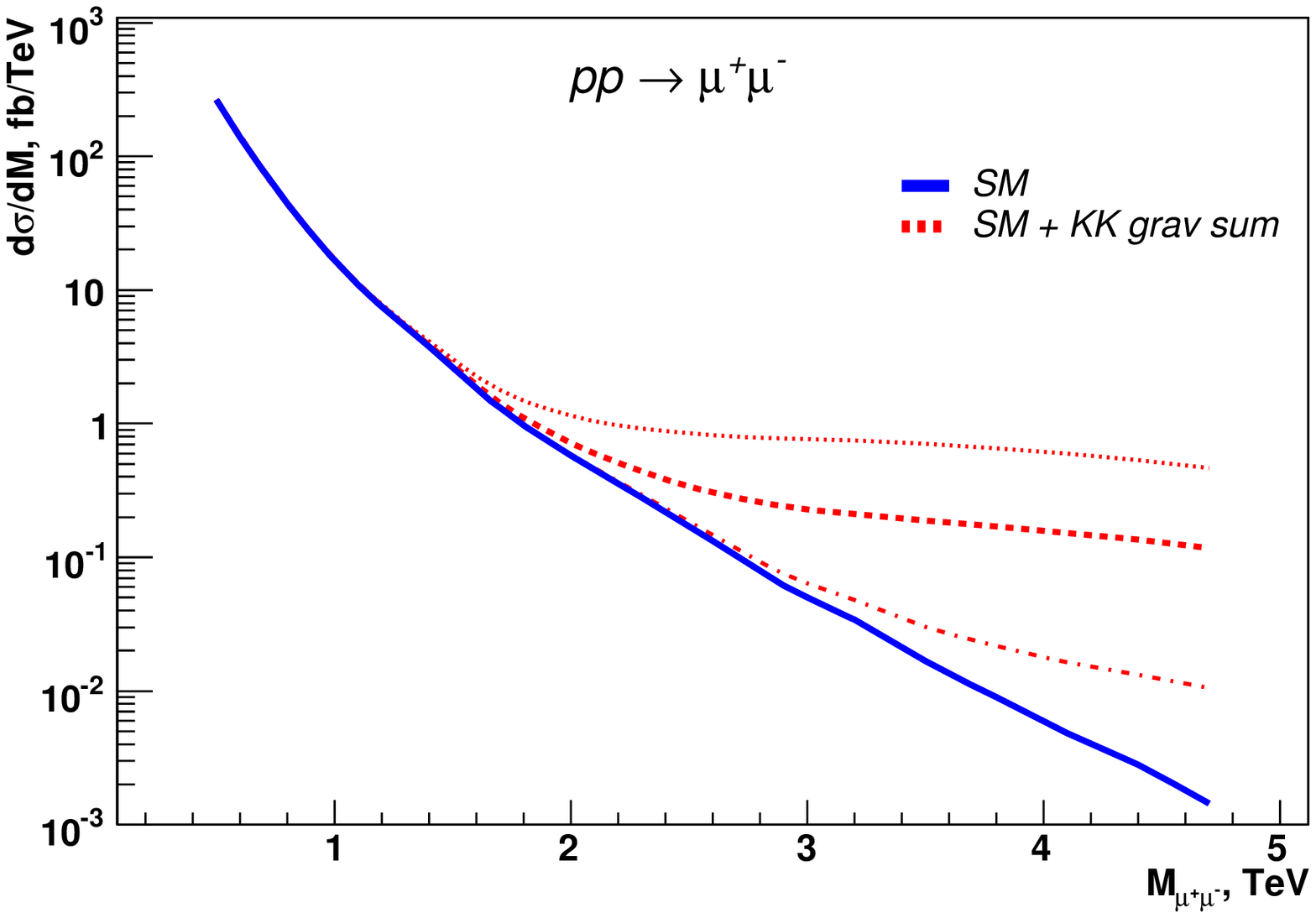}
\caption{Dilepton invariant mass distribution for  parameter
$\frac{0.91}{\Lambda_{\pi}^2 m_{1}^2}\times {TeV^4}$=0.0014 (dashed-dotted line), 0.0046
(dashed line), 0.01 (dotted line) for the LHC} \label{KKcontrib2}
\end{minipage}
\begin{minipage}[t]{.49\linewidth}
\centering
\includegraphics[width=8.5cm,height=8.5cm]{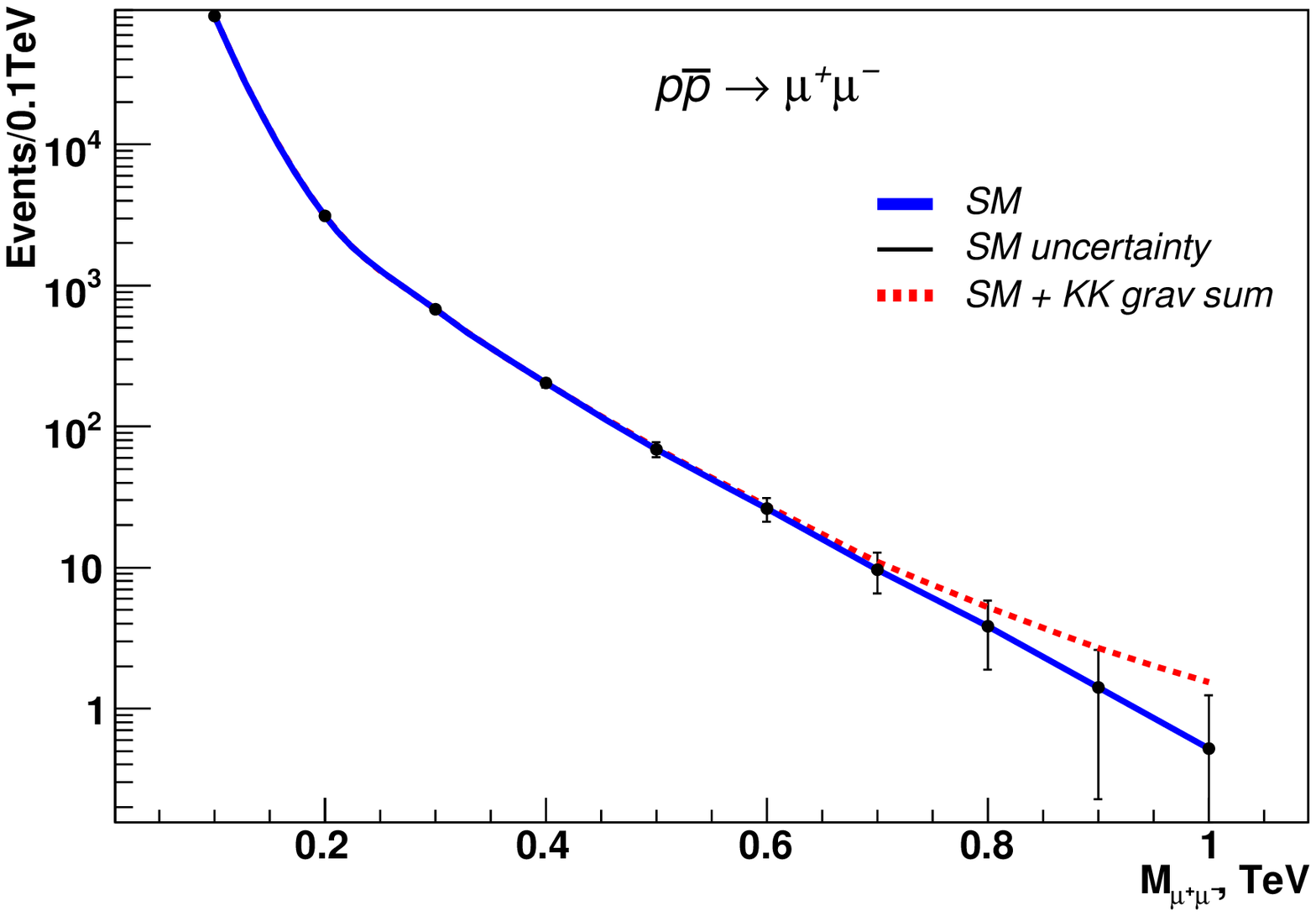}
\caption{Dilepton invariant mass distribution for  $95\%$ CL parameter
$\frac{0.91}{\Lambda_{\pi}^2 m_{1}^2}\times {TeV^4}$=0.66 for the Tevatron ($L=10{fb}^{-1}$)}
\label{KKclTEV}
\end{minipage}
\begin{minipage}[t]{.49\linewidth}
\centering
\includegraphics[width=8.5cm,height=8.5cm]{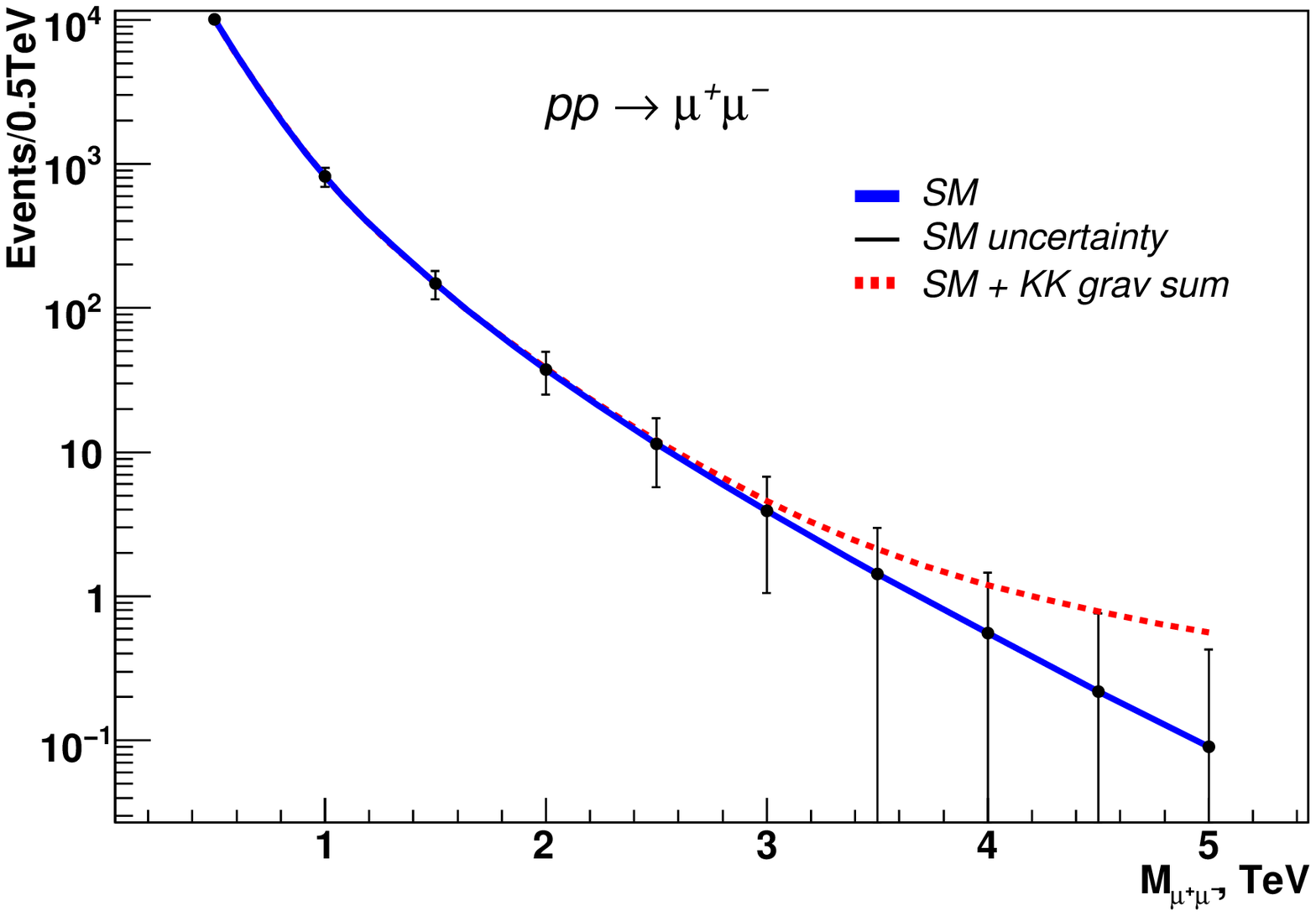}
\caption{Dilepton invariant mass distribution for  $95\%$ CL parameter
$\frac{0.91}{\Lambda_{\pi}^2 m_{1}^2}\times {TeV^4}$=0.0014
for the LHC ($L=100{fb}^{-1}$)}
\label{KKclLHC}
\end{minipage}
\end{figure*}

\begin{figure*}
\begin{minipage}[t]{.49\linewidth}
\centering
\includegraphics[width=8.5cm,height=8.5cm]{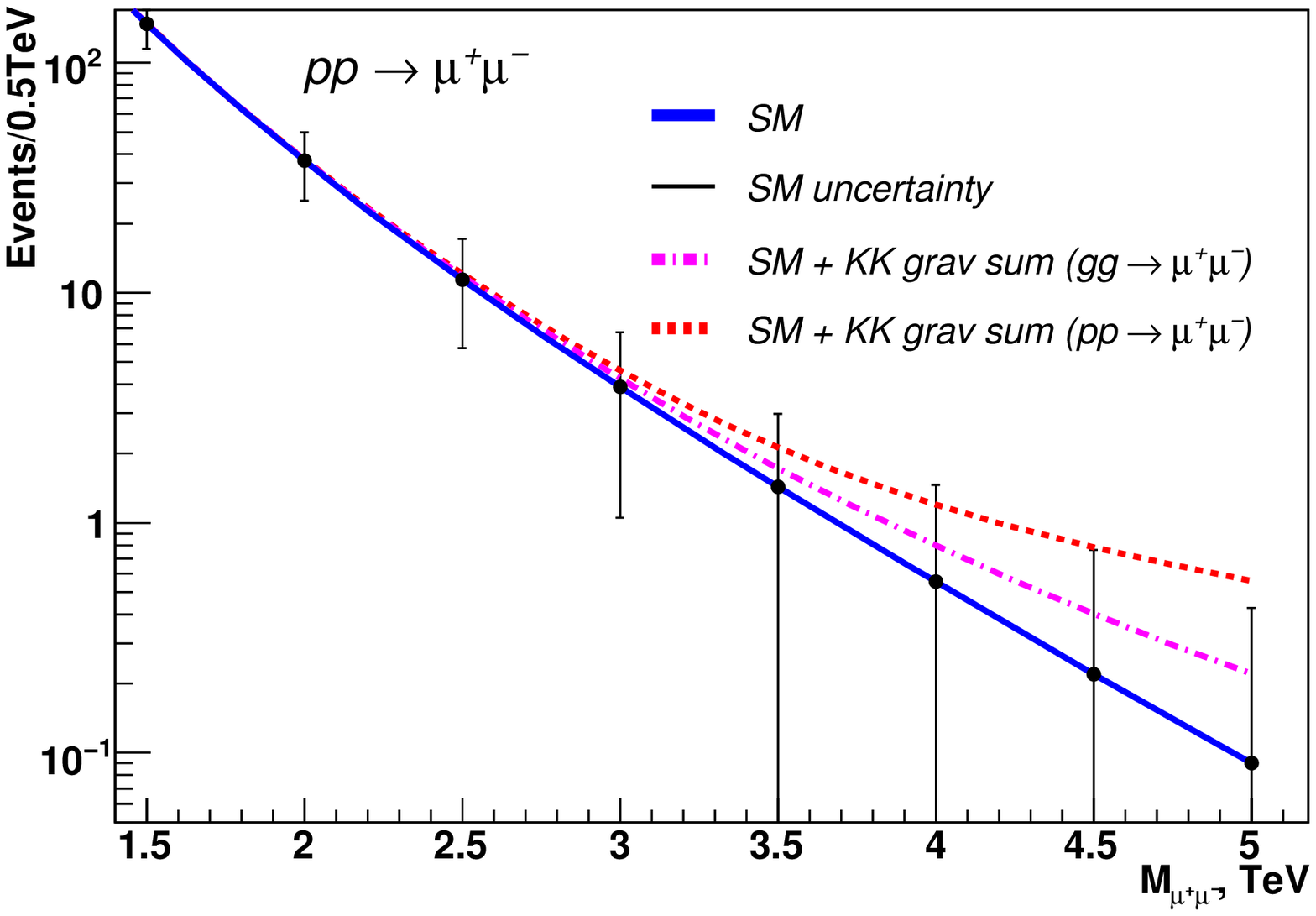}
\caption{Dilepton invariant mass distribution for  $95\%$ CL
coupling parameter for the LHC. The dashed-dotted line corresponds
to SM + KK  sum ($gg\to  \mu^{+}\mu^{-}$), the dashed line
corresponds to SM + KK  sum ($pp\to  \mu^{+}\mu^{-}$) process.}
\label{KKggmm}
\end{minipage}
\begin{minipage}[t]{.49\linewidth}
\centering
\includegraphics[width=8.5cm,height=8.5cm]{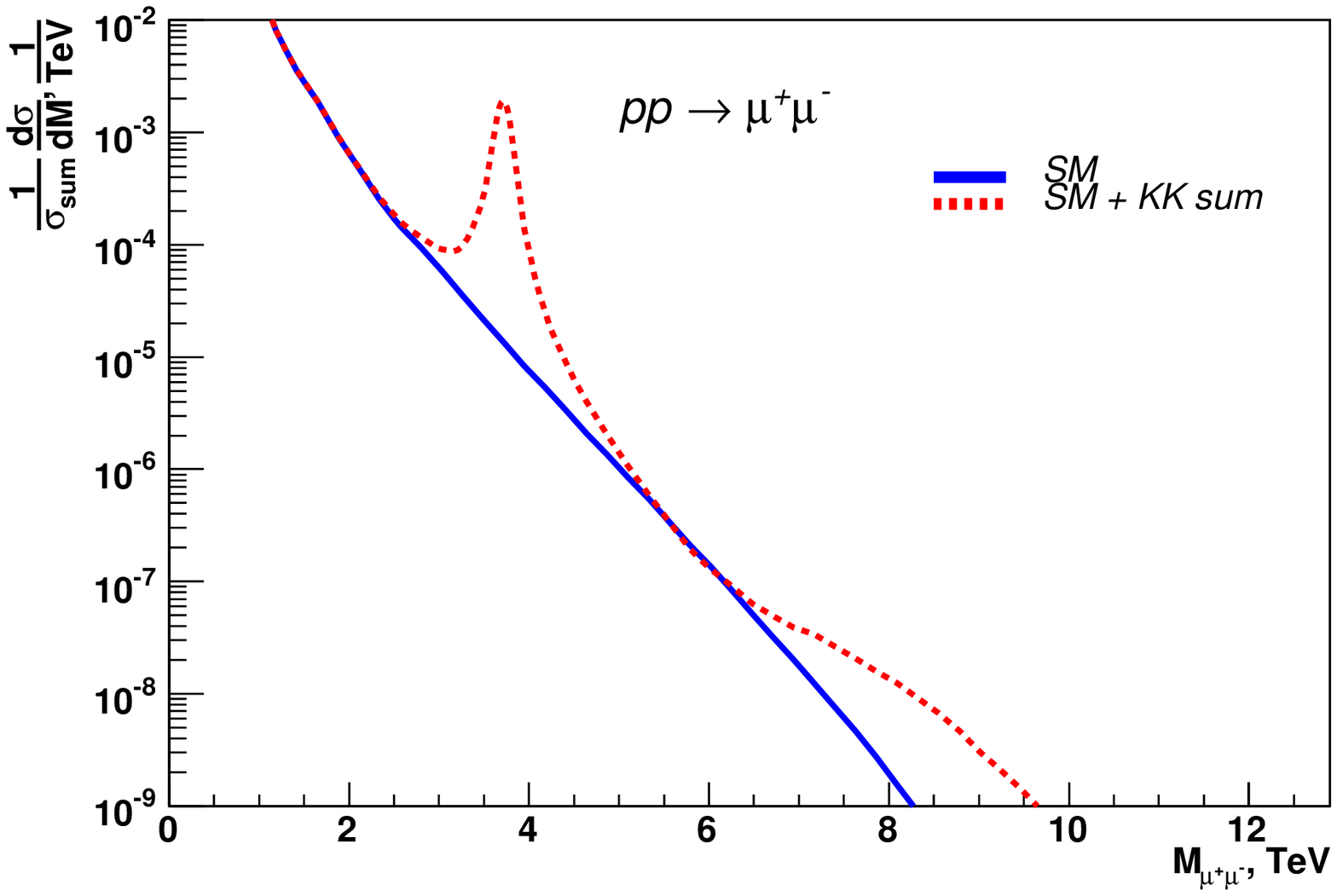}
\caption{Dilepton invariant mass distribution from the SM (solid
line) and from the SM plus sum of KK modes including the first KK
resonance with
$M_{res}=3.83~TeV$,~$\Gamma_{res}=0.08~TeV$,~$\Lambda_{\pi}=8~TeV$
(dashed line) for the LHC} \label{KKressm}
\end{minipage}
\begin{minipage}[t]{.49\linewidth}
\centering
\includegraphics[width=8.5cm,height=8.5cm]{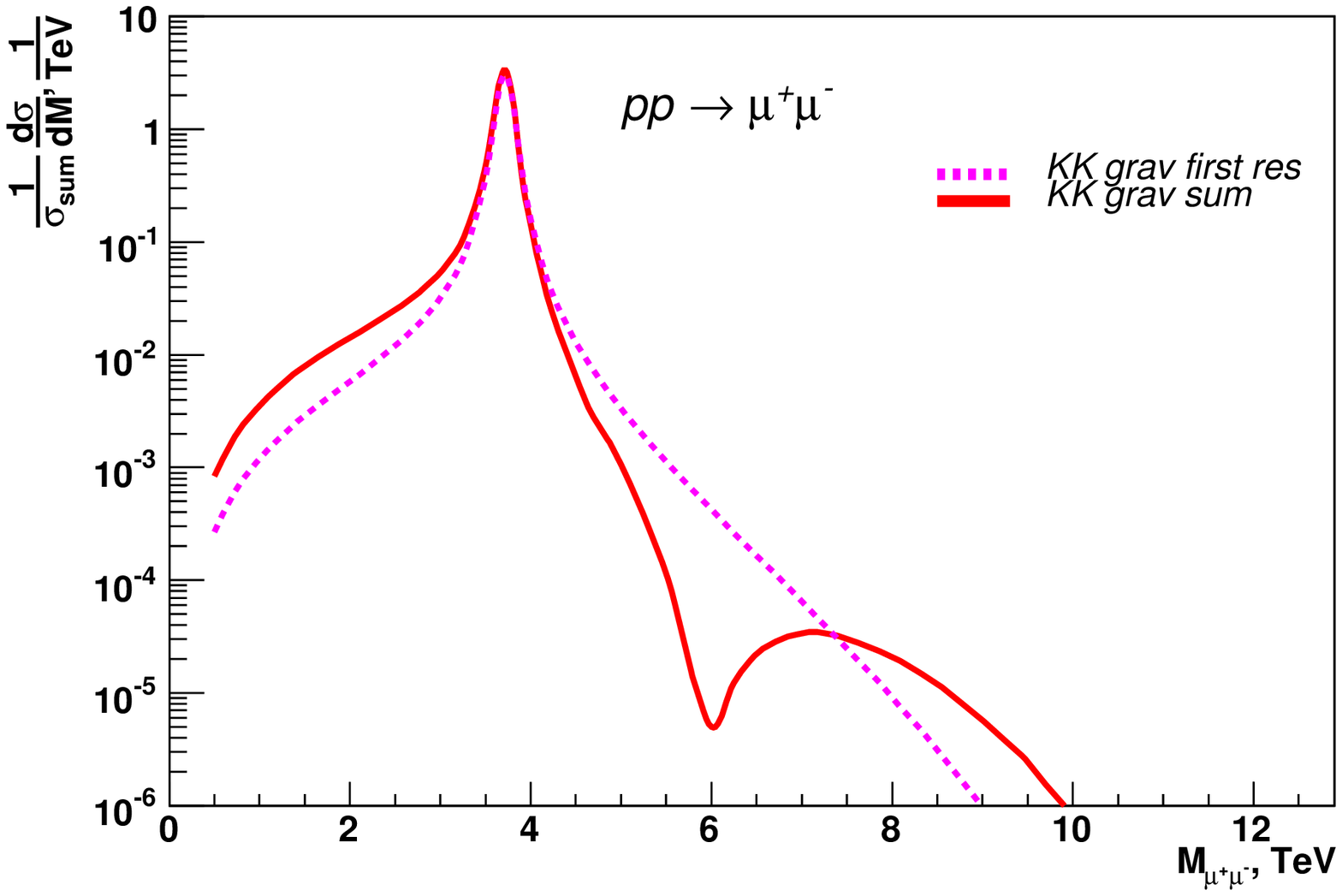}
\caption{The normalized dilepton invariant mass distribution from
the first KK resonance plus the sum of KK tower states starting
from the second mode (solid line) and from the first KK resonance
only (dashed line) for
$M_{res}=3.83~TeV$,~$\Gamma_{res}=0.08~TeV$,~$\Lambda_{\pi}=8~TeV$~for
the LHC} \label{KKres1}
\end{minipage}
\begin{minipage}[t]{.49\linewidth}
\centering
\includegraphics[width=8.5cm,height=8.5cm]{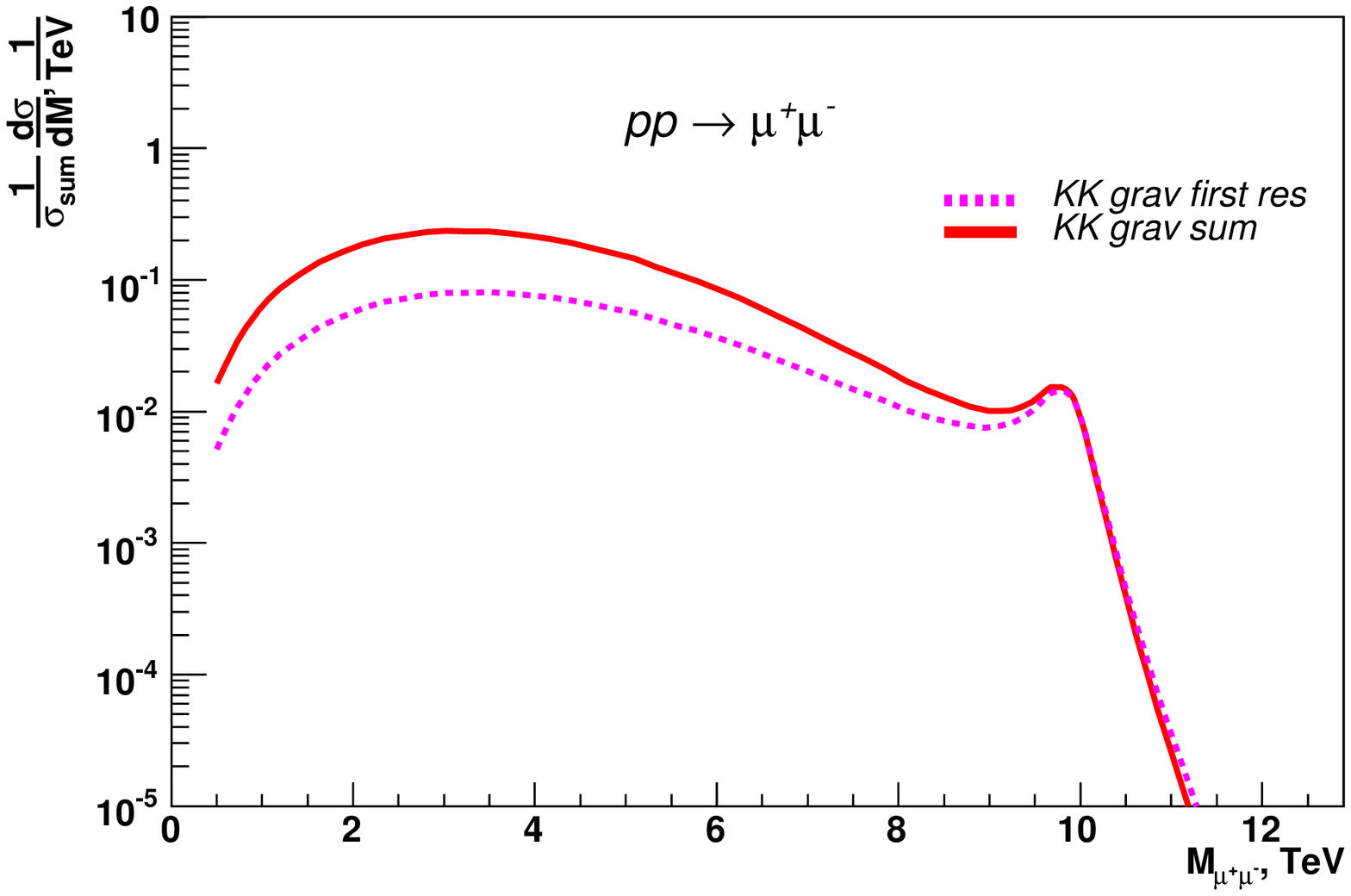}
\caption{The normalized dilepton invariant mass distribution from
the sum of  KK tower states starting from the first KK mode (solid
line) and from the first KK mode only (dashed line) for
$M_{res}=10~TeV$,~$\Gamma_{res}=0.5~TeV$,~$\Lambda_{\pi}=14~TeV$~for
the LHC} \label{KKres2}
\end{minipage}
\end{figure*}

Qualitatively, the situation from the phenomenological point of
view is similar to that appearing in the ADD scenario and worked
out by J.~Hewett in \cite{Hewett:1998sn}. The correspondence
between the parameters used in our study and in
\cite{Hewett:1998sn} is the following:
$$\frac{\Lambda_\pi^2
m_1^2}{0.91}=\frac{1}{4}\frac{M_s^4}{\lambda}.
$$
 As shown for the RS1 model in \cite{Davoudiasl:1999jd} the exchange of a tower of the KK
gravitons in the energy range below the KK production threshold,
similar to the ADD case, leads to an increase of the invariant mass tail
of produced
particles. For the Drell-Yan process it is
demonstrated in  Figs.~\ref{KKcontrib1},~\ref{KKcontrib2}.
The process $gg\to l^{+}l^{-}$ contributes to the Drell-Yan process
and it was included in our numerical simulations.
As shown on Fig~\ref{KKggmm} this contribution is very significant for the LHC.
As will be demonstrated below, even in the case when the first KK
resonance lies in the energy range  accessible for a detection one
should take into account the contribution from all the other KK
states. One should stress that in the ADD scenario, in addition to
deviations from the SM prediction for the processes such as lepton
pair production, there should also exist  processes with the KK
tower radiation off. The latter processes do not take place in the
RS model.

Using the standard ${\chi}^2$ analysis and taking into account the
expectations for systematic uncertainties (detector smearing,
electroweak, QCD scale, Parton Distribution Function(PDF)) and statistical uncertainties of the SM
dilepton invariant mass shape (see experimental data
\cite{Abazov:2008as} for the Tevatron and  Monte Carlo simulations
\cite{Ball:2007zza} for the LHC), we obtain the current Tevatron
limit for  the coupling parameter at $95\%$ CL and  estimate
expected experimental limits for this parameter
(Table~\ref{tab_cl}) that may be reached at the Tevatron for
higher luminosities and for various luminosities at the LHC. We
used CompHEP for the calculations of the SM center values in
thought experiment.

%=======================================================================
\begin{table*}[htb]
%\begin{center}
\caption{Experimental limits for the coupling parameter at $95\%$
CL that may be reached at the Tevatron and the LHC using Drell-Yan
process for some values of integrated luminosity $L$.}
\label{tab_cl}
\newcommand{\m}{\hphantom{$-$}}
\newcommand{\cc}[1]{\multicolumn{1}{c}{#1}}
\renewcommand{\tabcolsep}{1.4 pc} % enlarge column spacing
\renewcommand{\arraystretch}{1.2} % enlarge line spacing
\begin{tabular}{cc|cc}
\hline
\multicolumn{2}{c}{TEVATRON ($\sqrt{s}=1.96~TeV$)} & \multicolumn{2}{c}{LHC ($\sqrt{s}=14~TeV$)}\\
$L$, $fb^{-1}$ & $\frac{0.91}{\Lambda_{\pi}^2
m_{1}^2}~at~95\%~CL$, $TeV^{-4}$ &
$L$, $fb^{-1}$ & $\frac{0.91}{\Lambda_{\pi}^2 m_{1}^2}~at~95\%~CL$, $TeV^{-4}$\\
%L $fb^{-1}$ & k $TeV^{-4}$ & L & k\\
\hline
1 & 1.185 & 10 & 0.238 $\cdot 10^{-2}$\\
%\hline
2 & 0.995 & 20 & 0.203 $\cdot 10^{-2}$\\
%\hline
3 & 0.900 & 30 & 0.184 $\cdot10^{-2}$\\
%\hline
5 & 0.790 & 50 & 0.164 $\cdot10^{-2}$\\
%\hline
10 & 0.664 & 100 & 0.140 $\cdot10^{-2}$\\
\hline
\end{tabular}
%\end{center}
\end{table*}
%=======================================================================

The Tevarton limit for $1fb^{-1}$ of integrated luminosity
expressed in terms of parameter $M^{GRW}_s$ introduced in
\cite{Giudice:1998ck}
$$M^{GRW}_s=\left(\frac{1}{2\pi}\cdot\frac{0.91}{\Lambda_\pi^2
m_1^2}\right)^{-\frac{1}{4}}$$
 gives
$M^{GRW}_s(1fb^{-1})=1.52~TeV$, which is in a good agreement with
the corresponding limit from the cited experimental paper
\cite{Abazov:2008as}.

The last string of Table~\ref{tab_cl} contains limits
corresponding to the highest value of collider luminosity:
\begin{equation}\label{KKcl}
Tevatron(10 fb^{-1}):~\frac{0.91}{\Lambda_\pi^2 m_1^2}\times
{TeV^4} < 0.66 , \quad  LHC(100
fb^{-1}):~\frac{0.91}{\Lambda_\pi^2 m_1^2}\times {TeV^4} < 0.0014.
\end{equation}
Figures \ref{KKclTEV} and \ref{KKclLHC} demonstrate distributions
corresponding to values (\ref{KKcl}). These limits may be used for
estimating the lowest value of parameter $\Lambda_\pi$ from a
requirement that the width of a resonance be smaller than its
mass: $\Gamma_1 < m_1/\xi$, where $\xi$ is some number, $\xi>1$.
Using limits (\ref{KKcl}) and the equation for the total graviton
width (\ref{totwidth}) $\frac{m_{1}^3}{\Lambda_{\pi}^{2}\cdot
4\pi}\frac{97}{80} < \frac{m_1}{\xi}$, we get
\begin{equation}\label{lambdapi}
Tevatron:~\Lambda_\pi > 0.61 \cdot \xi^{1/4}\,TeV, \quad
LHC:~\Lambda_\pi > 2.82 \cdot \xi^{1/4}\,TeV,\quad \xi>1.
\end{equation}


One of the effects in searches for KK
resonances below the production threshold of the first state is an
enhancement of the effective coupling due to KK summation in
comparison to the first mode contribution below the threshold
only. For the  considered case of the stabilized RS model one has a factor
 $$\sum_{n \neq 0}\frac{ (\psi^{(n)}(L))^2}{m_n^2} \approx 1.82
\frac{ (\psi^{(1)}(L))^2}{m_1^2}\,.$$ This leads to an increase by
$1.82^2 \approx 3.3$ times in the production rate (for the case of
one flat extra dimension this factor is $1.64
=>\sum\frac{1}{n^{2}}=\pi^2/6$ being numerically close to the
warped case).

To illustrate changes in distributions due to KK tower
contributions we run simulations for two parameter points with the
first KK resonance being in and out of directly detectable
regions. The first point ($m_1=3.83~TeV$, $\Lambda_\pi=8~TeV$,
$\Gamma_1=0.08~TeV$) was already discussed in Section 3. Such an
RS resonance (see Fig.~\ref{KKressm}) is close to the direct reach
limits expected for the LHC \cite{Davoudiasl:1999jd}. For the
second point ($m_1=10~TeV$, $\Lambda_\pi=14~TeV$,
$\Gamma_1=0.5~TeV$) the mass of the first KK excitation is close
to the collider energy limit, and it is  not directly observable. For
both points we can use the low-energy effective Lagrangian
approach. The effective Lagrangian allows us in both cases to sum
up the contributions from all the KK modes or from all except the
first one, and in this way to take into account their influence on
the background tail. As one can see from Fig.~\ref{KKres1} and
Fig.~\ref{KKres2}, the additional substrate from the KK tower
increases the production rate more than 3 times in the invariant
mass region below the resonance mass. The situation is
significantly different above the resonance, where in addition to
the resonance pike there is an area with a minimum due to a
destructive interference between the first KK resonance and the
remaining KK tower contribution. This local minimum takes place at
the value of invariant mass $M_{min} \approx 1.5 m_1$. The growth
of the invariant mass after the minimum is strongly suppressed by
parton distribution functions leading to an additional bump in the
invariant mass shape. But this bump is unlikely to be visible in
the experiment on top of the SM background as shown in
Fig.\ref{KKressm}.

In conclusion of this part, one should stress that in order to perform
correct searches for KK resonances not only interferences with the
SM if nonvanishing and computed NLO QCD corrections \cite{Mathews}
should be included into corresponding generators, but also the influence of
those KK states, which are not reachable directly.

\section{Conclusion}
In the present paper we have continued studies of the effects
that appear in theories with extra space dimensions due to the
exchange of directly unaccessible KK modes of the fields
propagating in the bulk
 when the fundamental energy scale that defines these masses is  larger than
the typical collision energies $1 - 10~TeV$. We have derived the
effective Lagrangian resulting from the exchange of these KK modes
in the stabilized RS model. Such a Lagrangian has a simple structure
of a product of two currents, corresponding to the zero mode of a
bulk field, multiplied by an effective coupling constant of a
dimension depending on the spin of the bulk field. The exact value
of this coupling is model dependent and has to be calculated for
each model separately. In fact, this structure is reminiscent of
the Fermi interaction. Obviously, if the bulk field is the gravity
field, then the corresponding current is just the standard
energy-momentum tensor of the SM fields (see Appendix A). A
delicate nuance in the case of the stabilized RS1 model is that
the multidimensional metric has both tensor and scalar massive
degrees of freedom, the latter coupled to the trace of the 
energy-momentum tensor or to its  anomaly in the case of massless SM
fields. It is shown explicitly that the contributions of the
scalar modes are much smaller than those of the tensor modes.

The Feynman rules  for the new effective 4-particle
interaction  vertices of the Standard model fields
 have been incorporated into the CompHEP computer program,
and explicit formulas for the differential and total cross
sections for a number of processes, generated by these effective
Lagrangians, are presented in  Appendix B.
For completeness we have included into
the symbolic formulas the final state particle masses and
contributions of virtual graviton and radion KK modes.

Using these results, we have calculated and plotted the
contributions due to the multidimensional gravity to  the
Drell-Yan production processes for the Tevatron and LHC energies
for a number of the stabilized RS model parameter points.
 It was clearly demonstrated  that an
enhancement of the effective coupling due to KK summation in
comparison to the first mode contribution only leads to an
increase of the collider potential to probe the first mode mass
below the threshold of its production, the mass being
significantly larger than the collision energies. In the case, where
the first mode mass is in the accessible energy range, but all the
other modes are out of this range, summation of all the modes
contributions starting from the second one leads to a significant
change of the shape of the Breit-Wigner distribution.
 The latter case occurs for the considered explicit example in the
stabilized Randall-Sundrum model.

We have calculated the effective  couplings for this case. To this
end, we have used both analytical results and numerical estimates
for the wave functions of tensor and scalar modes and performed an
approximate summation of the series of inverse mass squared of the
KK excitations. For a choice of the parameters of the model, where
the effective energy scale $M\sim 1~TeV$ and the other parameters
are such that the metric of the stabilized RS model with the
radion mass of the order of $2~TeV$ is similar to that of the
unstabilized one, we found explicit formula (\ref{tensor_intCRS})
for the effective Lagrangian. For this particular choice of the
parameters the  first KK graviton  and the radion masses are
beyond the energy range directly accessible at the Tevatron and
within this range at the LHC. However the value of the effective
coupling in (\ref{tensor_intCRS}) is too small for deviations from
the SM tails to be observed and to probe the masses below the
production threshold at the Tevatron energies.  For the LHC the
first KK and the radion masses are within the directly accessible
region. In this case in order to perform correct searches for the
tensor resonances and to model the distribution tails one should
sum up the contributions from all the other KK modes and take into
account their interference with the resonances. It is  worth
pointing out that this summation of the contributions of the
massive tensor modes is also needed for modeling the background in
searches for the light radion, whose coupling to the SM fields
turns out to be larger, than in the case of the heavy radion.

\bigskip

{\large \bf Acknowledgments}

The authors are grateful to Prof.\,\fbox{P.F.~Ermolov} for useful
discussions and support. The work was supported by the 
Russian Ministry of Education and Science under Grant No. NS-1456.2008.2 and by
the RFBR Grant Nos. 08-02-91002-CERN$_-$a and 08-02-92499-CNRSL$_-$a.
M.S. acknowledges support of grant for young scientists under Grant No.
MK-5602.2008.2 of the President of Russian Federation. V.B. and
M.S. also acknowledge  support of grant of the "Dynasty"
Foundation.

\medskip

\section{Appendix A: Energy-momentum tensors for free SM fields}

The Lagrangian and the energy-momentum tensor for fermions:
\begin{equation}
L_{\Psi}=\frac{i}{2}\left(\bar\Psi
\gamma^{\mu}\partial_{\mu}\Psi-\partial_{\mu}\bar\Psi
\gamma^{\mu}\Psi\right)-m_{\Psi}\bar\Psi\Psi
\end{equation}
\begin{eqnarray}
T^{\Psi}_{\mu\nu}=\frac{i}{4}\left(\bar\Psi\gamma_{\mu}\partial_{\nu}\Psi+
\bar\Psi\gamma_{\nu}\partial_{\mu}\Psi-\partial_{\nu}\bar\Psi
\gamma_{\mu}\Psi-\partial_{\mu}\bar\Psi\gamma_{\nu}\Psi\right)-\\
\nonumber -\eta_{\mu\nu}\left(\frac{i}{2}\bar\Psi
\gamma^{\rho}\partial_{\rho}\Psi-\frac{i}{2}\partial_{\rho}\bar\Psi
\gamma^{\rho}\Psi-m_{\Psi}\bar\Psi\Psi\right)
\end{eqnarray}
\begin{equation}
{T^{\Psi}}^{\mu}_{\mu}=-\frac{3i}{2}\left(\bar\Psi
\gamma^{\mu}\partial_{\mu}\Psi-\partial_{\mu}\bar\Psi
\gamma^{\mu}\Psi\right)+4m_{\Psi}\bar\Psi\Psi
\end{equation}

The Lagrangian and the energy-momentum tensor for massive vector
bosons (Z-boson):
\begin{equation}
L_{Z}=-\frac{1}{4}Z_{\mu\nu}Z^{\mu\nu}+\frac{m^{2}_{Z}}{2}Z^{\mu}Z_{\mu}
\end{equation}
\begin{equation}
T^{Z}_{\mu\nu}=-Z_{\mu\rho}Z_{\nu\sigma}g^{\rho\sigma}+m^{2}_{Z}Z_{\mu}
Z_{\nu}+\eta_{\mu\nu}\left(\frac{1}{4}Z_{\rho\sigma}Z^{\rho\sigma}-
\frac{m^{2}_{Z}}{2}Z^{\rho}Z_{\rho}\right)
\end{equation}
\begin{equation}
{T^{Z}}^{\mu}_{\mu}=-m^{2}_{Z}Z^{\mu}Z_{\mu}
\end{equation}

The Lagrangian and the energy-momentum tensor for complex vector
bosons (W-bosons):
\begin{equation}
L_{W}=-\frac{1}{2}W^{+}_{\mu\nu}{W^{-}}^{\mu\nu}+m^{2}_{W}W^{+}_{\mu}{W^{-}}
^{\mu}\end{equation}
\begin{eqnarray}
T^{W}_{\mu\nu}=-W^{+}_{\mu\rho}W^{-}_{\nu\sigma}g^{\rho\sigma}-
W^{+}_{\nu\rho}W^{-}_{\mu\sigma}g^{\rho\sigma}+m^{2}_{W}
\left(W^{+}_{\mu}W^{-}_{\nu}+W^{+}_{\nu}W^{-}_{\mu}\right)
+\\ \nonumber
+\eta_{\mu\nu}\left(\frac{1}{2}W^{+}_{\rho\sigma}{W^{-}}^{\rho\sigma}-
m^{2}_{W}W^{+}_{\rho}{W^{-}}^{\rho}\right)
\end{eqnarray}
\begin{equation}
{T^{W}}^{\mu}_{\mu}=-2m^{2}_{W}W^{+}_{\mu}{W^{-}}^{\mu}
\end{equation}

The Lagrangian and the energy-momentum tensor for massless vector
bosons (the photon and the gluons):
\begin{equation}
L_{A}=-\frac{1}{4}F_{\mu\nu}F^{\mu\nu}
\end{equation}
\begin{equation}
T^{A}_{\mu\nu}=-F_{\mu\rho}F_{\nu\sigma}g^{\rho\sigma}
+\eta_{\mu\nu}\frac{1}{4}F_{\rho\sigma}F^{\rho\sigma}
\end{equation}
\begin{equation}
{T^{A}}^{\mu}_{\mu}=0
\end{equation}

The Lagrangian and the energy-momentum tensor for the scalar field
(the Higgs field in the unitary gauge):
\begin{equation}
L_{\Phi}=\frac{1}{2}\partial^{\mu}\Phi\partial_{\mu}\Phi-
\frac{m^{2}_{\Phi}}{2}\Phi^{2}
\end{equation}
\begin{equation}
T^{\Phi}_{\mu\nu}=\partial_{\mu}\Phi\partial_{\nu}\Phi-
\eta_{\mu\nu}\left(\frac{1}{2}\partial^{\rho}\Phi\partial_{\rho}\Phi-
\frac{m^{2}_{\Phi}}{2}\Phi^{2}\right)
\end{equation}
\begin{equation}
{T^{\Phi}}^{\mu}_{\mu}=-\partial^{\mu}\Phi\partial_{\mu}\Phi+2m^{2}_{\Phi}
\Phi^{2}\end{equation}

\section{Appendix B: Partonic total and differential cross sections for $2
\to 2$ processes.}

Processes $p\bar{p}\to Z^0Z^0$:\\
% gg-ZZ

\begin{eqnarray}\label{ggzz}
\frac{d\hat{\sigma}_{gg\to ZZ}}{dz} =
\frac{\kappa^2}{16\pi\cdot16}\beta\left[ 1 + \beta^2(z^2 - 1) +
\frac{3}{16}\beta^4(1 - z^2)^2+
\right. \\
\left. 3\delta^2\left(\frac{b(g)}{2g}\right)^2\left(
1-\frac{2}{3}\beta^2+\beta^4\right)\right]\hat{s}^3 \nonumber
\end{eqnarray}

\begin{eqnarray}
\hat{\sigma}_{gg\to ZZ} = \frac{\kappa^2}{16\pi\cdot8}\beta\left[
1 - \frac{2}{3}\beta^2 + \frac{1}{10}\beta^4
~~+~~3\delta^2\left(\frac{b(g)}{2g}\right)^2
\left(1-\frac{2}{3}\beta^2+\beta^4\right)\right]\hat{s}^3
\end{eqnarray}

%uu-zz

\begin{align}
\frac{d\hat{\sigma}_{q\bar{q}\to ZZ}}{dz} = &
\frac{\kappa^2}{16\pi\cdot 96}\beta\left[ 4 - 2\beta^2(z^2 + 1) +
3\beta^4 z^2(1 - z^2)\right]\hat{s}^3 +\\ \nonumber &
\frac{\kappa\alpha\cdot c_1 \beta}{16\cdot 48}\left[\frac{\beta^2
z^3-z}{\xi_2} +
\beta z(2\beta z+\beta^2-3) +\xi_2^2\beta\cdot\frac{5-\beta^2}{\xi_1-z} + \frac{\beta^4-8\beta^2+11}{2}\right]\hat{s} + \\
\nonumber & \frac{\pi\cdot \alpha^2 c_2}{16\cdot 48}
\left[\frac{2\beta\xi_1}{\xi_1-z} -
\frac{\beta\xi_2^2}{{(\xi_1-z)}^2} + \frac{4\xi_2}{\xi_1^2-z^2} -
\beta \right]\frac{1}{\hat{s}} \nonumber &
\end{align}

\begin{align}\label{uuzz}
\hat{\sigma}_{q\bar{q}\to ZZ} = & \frac{\kappa^2}{16\pi\cdot
12}\beta\left[ 1 - \frac{2}{3}\beta^2 +
\frac{1}{10}\beta^4\right]\hat{s}^3 +\\ \nonumber &
\frac{\kappa\alpha\cdot c_1}{16\cdot
48}\left[(5-\beta^2)\frac{(1-\beta^2)^2}{2}\ln{\left|\frac{1+\beta}{1-\beta}\right|}
+ \beta^5 - \frac{20}{3}\beta^3 + 11\beta\right]\hat{s} + \\
\nonumber & \frac{\pi\cdot \alpha^2 c_2}{16\cdot24} \left[
\left(1+\beta^2+4\cdot\frac{1-\beta^2}{1+\beta^2}\right)\ln{\left|\frac{1+\beta}{1-\beta}\right|}-2\beta\right]\frac{1}{\hat{s}}
\nonumber &
\end{align} \\

Processes $p\bar{p}\to l^+l^-$: \\

 % gg-ll

\begin{eqnarray}
\frac{d\hat{\sigma}_{gg\to l^+l^-}}{dz} =
\frac{\kappa^2}{16\pi\cdot 64}\beta^3\left[ 2(1 - z^2) -
\beta^2(1-z^2)^2\right]\hat{s}^3
\end{eqnarray}

\begin{eqnarray}
\hat{\sigma}_{gg\to l^+l^-} = \frac{\kappa^2}{16\pi\cdot
24}\beta^3\left[ 1-\frac{2}{5}\beta^2\right]\hat{s}^3
\end{eqnarray}

% qq-ll

\begin{align}
\frac{d\hat{\sigma}_{q\bar{q} \to l^+l^-}}{dz} = & \frac{\kappa^2}{16\pi\cdot 32\cdot 3}\beta^3\left[(z^2 + 1) + 4\beta^2 z^2(z^2 - 1) \right]\hat{s}^3 +\\
\nonumber &
\frac{\kappa \alpha}{16\cdot 3} \left[ z\beta^2( 1 - \beta^2 + z^2\beta^2 )
\left( 2 Q_q - 2 v_l v_q \chi_1\right) + (1 - 3z^2)\beta^3 a_l a_q \chi_1\right]\hat{s} + \\ \nonumber &
\frac{\alpha^2 \pi}{2\cdot 3}\left[ (2\beta - \beta^3 + \beta^3 z^2)\left( Q_q^2 - 2 \chi_1 Q_q v_l v_q +
\chi_2(a_l^2 + v_l^2)(a_q^2 + v_q^2) \right) +
\right. \\ &
\left.
2\beta^2 z(-2\chi_1 Q_q + 4 \chi_2 v_l v_q) a_l a_q + 2 \chi_2 (a_l^2 + v_l^2) a_q^2 (\beta^3-\beta) \right]\frac{1}{\hat{s}}
\nonumber &
\end{align}

\begin{align}
\hat{\sigma}_{q\bar{q} \to l^+l^-} = & \frac{\kappa^2}{16\pi\cdot 36}\beta^3\left[ 1-\frac{2}{5}\beta^2\right]\hat{s}^3 +\\
\nonumber &
\frac{\alpha^2 \pi}{3}\left[ (2\beta - \beta^3 + \frac{1}{3}\beta^3)\left( Q_q^2 - 2 \chi_1 Q_q v_l v_q +
\chi_2(a_l^2 + v_l^2)(a_q^2 + v_q^2) \right) +
\right. \\ &
\left.
2 \chi_2 (a_l^2 + v_l^2) a_q^2 (\beta^3-\beta) \right]\frac{1}{\hat{s}}
\nonumber &
\end{align}\\

Processes for linear colliders: \\

% ee-qq

\begin{align}
\frac{d\hat{\sigma}_{e^+e^-\to q^+q^-}}{dz} &=
9\cdot\frac{d\hat{\sigma}_{q\bar{q}\to e^+e^-}}{dz}
\end{align}

%ee-gg

\begin{eqnarray}
\frac{d\hat{\sigma}_{e^+e^-\to gg}}{dz}=\frac{\kappa^2}{16\pi\cdot
2}(1 - z^4)\hat{s}^3
\end{eqnarray}

\begin{eqnarray}
\hat{\sigma}_{e^+e^-\to gg}=\frac{\kappa^2}{4\pi\cdot 5}\hat{s}^3
\end{eqnarray}

%AA-gg
\begin{eqnarray}
\frac{d\hat{\sigma}_{\gamma\gamma\to
gg}}{dz}=\frac{\kappa^2}{16\pi\cdot 2}(1 + 6z^2 + z^4)\hat{s}^3
\end{eqnarray}

\begin{eqnarray}
\hat{\sigma}_{\gamma\gamma\to gg}=\frac{\kappa^2}{\pi\cdot
5}\hat{s}^3
\end{eqnarray}

Total width for the KK graviton resonance\\

\begin{align}\label{totwidth}
\Gamma^{tot}_{n}= &\frac{M_{n}^3}{\Lambda_{\pi}^{2}\cdot 4\pi\cdot
5}\left[\frac{9}{4}
+\frac{1}{4}\left(\frac{13}{12}+\frac{14}{3}\frac{M_Z^2}{M_{n}^2}+4\frac{M_Z^4}{M_{n}^4}\right)\beta_Z
+\frac{2}{4}\left(\frac{13}{12}+\frac{14}{3}\frac{M_W^2}{M_{n}^2}+4\frac{M_W^4}{M_{n}^4}\right)\beta_W
\right. \\ \nonumber & \left. +\frac{21}{8}
+\left(\frac{3}{8}-\frac{1}{2}\frac{M_t^2}{M_{n}^2}-4\frac{M_t^4}{M_{n}^4}\right)\beta_t
\right]\approx\frac{M_{n}^3}{\Lambda_{\pi}^{2}\cdot
4\pi}\frac{97}{80}
\end{align}

%\\
where:
\begin{align}
\nonumber & \hat{s}=x_1 x_2 s,\\ \nonumber & \beta_f =
\sqrt{1-\frac{4M^2_f}{\hat{s}}},~M_f~standing~for~the~mass~of~the~final~state~particle,\\
\nonumber & z=\cos\theta,~\theta~denoting~the~
partonic~center~of~mass~scattering~angle,\\ \nonumber &
\theta_W~being~the~Weinberg~angle,\\ \nonumber &
%k=\frac{C}{M^4},~with~C~being~a~dimensionles~parameter~and~M~being~a~mass~dimension~parameter,
\kappa=\frac{0.91}{\Lambda_{\pi}^{2}m_{1}^{2}},
~with~\Lambda_\pi~being~a~coupling~constant~of~the~first~KK~resonance~and~its~mass~m_1,
\\ \nonumber & a_l = - 1,\\ \nonumber &
v_l = - 1 + 4\sin^2\theta_W,\\ \nonumber & a_q = 2T_{3q},\\
\nonumber & v_q =2T_{3q} - 4Q_{q} sin^2\theta_W,\\ \nonumber &
Q_q~denoting~the~charge~of~the~quark,\\ \nonumber &
T_{3q}=\pm\frac{1}{2}~for~up/down-type~quarks,\\ \nonumber & \xi_1
= \frac{1+\beta^2}{2\beta},~~\xi_2 = \frac{1-\beta^2}{2\beta},\\
\nonumber & c_1 = \frac{a_q^2 +
v_q^2}{\sin^2\theta_W\cos^2\theta_W},~~ c_2 = \frac{6 a_q^2 v_q^2+
a_q^4 + v_q^4}{\sin^4\theta_W\cos^4\theta_W}\\ \nonumber & \chi_1
=
\frac{1}{16\sin^2\theta_W\cos^2\theta_W}\cdot\frac{s(s-M_Z^2)}{(s-M_Z^2)^2+M_Z^2\Gamma_Z^2},\\
\nonumber & \chi_2 = \frac{1}{256
\sin^4\theta_W\cos^4\theta_W}\cdot\frac{s^2}{(s-M_Z^2)^2+M_Z^2\Gamma_Z^2}\,.
\nonumber
\end{align}

\end{document}